# Cascaded Metasurface Interferometer for Multipath Interference with Classical and Quantum Light

Rebecca Aschwanden[1], Nicolás Claro-Rodríguez[1], Ruizhe Zhao[2], Patricia Kallert[1], Tobias Krieger[3], Quirin Buchinger[4], Saimon F. Covre da Silva[3], Sandra Stroj[5], Michele Rota[6], Sven Höfling[4], Tobias Huber-Loyola[4,7], Armando Rastelli[3], Rinaldo Trotta[6], Lingling Huang[2], Tim Bartley[1], Klaus D. Jöns[1], and Thomas Zentgraf*[1]

[1] Institute for Photonic Quantum Systems (PhoQS), Center for Optoelectronics and Photonics Paderborn (CeOPP) and Department of Physics, Paderborn University, Warburger Straße 100, 33098, Paderborn, Germany
[2] Beijing Engineering Research Center of Mixed Reality and Advanced Display, School of Optics and Photonics, Beijing Institute of Technology, Beijing 100081, China
[3] Institute of Semiconductor and Solid State Physics, Johannes Kepler University, Altenbergerstraße 69, Linz 4040, Austria
[4] Technische Physik, Julius-Maximilians-University of Würzburg, Am Hubland, 97074 Würzburg, Germany
[5] Research Center for Microtechnology, Vorarlberg University of Applied Sciences, Campus V, Hochschulstrasse 1, 6850, Dornbirn, Austria
[6] Dipartimento di Fisica, Sapienza Università di Roma, Piazzale Aldo Moro 5, 00185 Roma, Italy
[7] Institute of Photonics and Quantum Electronics (IPQ) and Center for Integrated Quantum Science and Technology (IQST), Karlsruhe Institute of Technology, Engesserstr. 5, 76131 Karlsruhe, Germany

*Email: thomas.zentgraf@uni-paderborn.de

**Abstract**

Beamsplitters represent fundamental components in both classical and quantum optical systems, enabling the distribution of light, as well as the generation of interference, superposition and entanglement. However, optical networks constructed from conventional bulk 2×2-beamsplitters encounter inherent scalability issues, as the number of required beamsplitters scales quadratically with the number of optical modes for a fully connected network. Metasurfaces offer a promising route to overcome these constraints. By manipulating light at the wavelength scale compact optical components with advanced functionalities can be constructed, which address several modes simultaneously. In this work, we design and experimentally utilize a metasurface as a multiport beamsplitter. Furthermore, we realize a multimode interferometer composed of two cascaded metasurfaces. We characterize the individual and cascaded metasurfaces using classical light, showing controllable splitting ratios through tunable phase relations. We then expand the approach to quantum light, employing single photons to demonstrate second- and third-order photon correlations, as well as single photon interference across multiple spatial paths. These results establish metasurface-based multiport beamsplitters as a scalable and reconfigurable platform bridging classical and quantum photonics.

## Introduction

Beamsplitters constitute fundamental building blocks in both classical and quantum optics. In classical optics, they redistribute optical power between two paths, forming the basis of interferometry and optical signal processing. Conventional cube beamsplitters, typically composed of two prisms, support two spatially orthogonal input and output channels but are bulky and difficult to scale into dense or integrated architectures. More complex optical networks are constructed from cascaded 2×2-beamsplitters and phase shifters; however, the physical implementation becomes increasingly demanding, as the number of required components scales quadratically with the number of optical modes.[1] In quantum optics, beamsplitters play an even more fundamental role. They enable quantum interference between indistinguishable photons and serve as the primary mechanism for generating and manipulating photonic entangled states in the spatial and polarisation degree of freedom. Entangled states serve as a core resource for quantum communication, information, and computation.[2–4] Conventional two-port beamsplitters are limited to creating interference between only two modes, restricting access to richer multimode quantum phenomena and higher-dimensional state space. This limitation constrains the implementation of complex quantum operations, which rely on coherence among multiple modes.[5] Consequently, advancing both classical and quantum photonic technologies calls for miniaturized, multiport beamsplitters capable of handling a larger number of input and output channels within compact architectures. Despite extensive research,[6,7] the miniaturization of traditional cube beamsplitters remains challenging. Various strategies have been explored in order to achieve flat-type beamsplitters, with each strategy exhibiting its own strengths and limitations. Traditional approaches such as multimode interference couplers, topological waveguides, or directional couplers rely on established waveguiding principles but are typically constrained in scalability, optical bandwidth, and polarization control.[8,9] Other concepts, including photonic crystals and inverse-designed devices, offer alternative solutions but frequently involve compromises in design complexity or fabrication challenges.[10] Among these approaches, metasurfaces provide a powerful platform to realize multiport beamsplitting functionality in a compact planar geometry with versatile design possibilities. By tailoring the response of subwavelength scatterers (meta-atoms), metasurfaces enable abrupt spatial phase modulation over distances comparable to the optical wavelength, circumventing the need for large propagation distances necessary for traditional bulky optics.[11] This capability has been widely exploited in classical optics for the realization of metalenses[12–14] or metaholograms,[15–17] and has been extended to quantum optics.[18–24] Metasurface-based beamsplitters have been realized using both dielectric and plasmonic materials, with and without polarization selectivity, demonstrating different numbers and spatial configurations of output beams. Corresponding research on such devices typically focused on only one input configuration effectively leading to a $1 \times N$-beamsplitter.[25–28] While these studies have shown versatile beamsplitting functionalities, they remain limited to single-device implementations. A few studies have explored the coherent interaction of two input beams,[18,29,30] yet systematic cascading of multiple metasurfaces to realize complex interferometric operation remains largely unexplored. In this work, we present an experimental realization of a free-space multiport metasurface-based interferometer. Our design consists of two identical amorphous silicon-based metasurfaces, where each one is used as a multiport beamsplitter. The first metasurface divides the incident field into well-designed diffraction orders, that are recombined on the second metasurface under controlled phase conditions, forming a compact multipath interferometric network. By adjusting the relative phases between the different light paths, we achieve deterministic control of the output intensities, effectively realizing a reconfigurable multiport interferometer. We begin by characterizing the individual metasurfaces using classical light to study their diffraction efficiencies and output distributions. We then demonstrate the phase-dependent interference of the cascaded configuration. Finally, we extend our study to the quantum regime by replacing the classical source with a semiconductor quantum dot emitting single photons "on demand". We find that the metasurface architecture preserves the quantum statistics and enables both second- and third-order photon correlation measurements within a single device. Together, these results mark an important step toward scalable, integrated metasurface-based interferometry for quantum and classical photonic applications.

## Results and Discussion

**Metasurface Design**

The multiport beamsplitting functionality of our metasurface is schematically shown in Fig. 1a. The input $in_1^0$ (*in* stands for input, the subscript 1 refers to the first metasurface, and the superscript 0 indicates the input channel) is diffracted into three output ports (out$^{-1}$, out$^{+1}$ and out$^{+2}$) along the x-direction that correspond to the -1, +1, and +2 diffraction order. With our design, we target equal intensities $I_{\text{out}^m} = \frac{1}{3}$ in all of these three output paths, where $m$ is the number of the diffraction order.

The metasurface's spatial phase profile $\Phi(x)$ is found by employing a gradient-descent optimization, which minimizes the error function $e_m$

$$e_m = \sum_{m=1}^{3} \text{abs}\left(I_{\text{out}^m} - \frac{1}{3}\right), \quad (1)$$

where $I_{\text{out}^m}$ refers to the calculated intensities of output path $m$. Each intensity $I_{\text{out}^m}$ is calculated based on the square of the Fourier coefficients $C_x^m$, expressed as:

$$C_{x_p}^m = \frac{1}{2\pi} \sum_{x_p}^{D} e^{i\Phi(x)} e^{i\frac{2\pi m x_p}{D}} \quad (2)$$

where $x_p$ represents the coordinate of the meta-atoms in the periodic supercell. The coordinate $x_p$ is discretized into $n$=22 pixels each being $p$ = 400 nm in size, giving a total supercell length $D = n \cdot p$ = 8.8 μm. The resulting diffraction angle is $\theta = \arcsin(\frac{\lambda}{D}) = 5°$. During optimization the spatial phase profile $\Phi(x)$ is iteratively varied until the minimum value of the error function $e_m$ is found, meaning the best target intensity distribution is achieved. After convergence of the gradient-descent algorithm, the optimized discretized spatial phase profile yields diffraction efficiencies of $I_{-1,+1,+2} = 0.28$ for the three desired output paths and a small diffraction efficiency of 0.04 is observed in output path $I_{\text{out}}^0$ (Fig. 1b).

The generated spatial phase profile is displayed in Fig. 1c. Physically, this phase profile is realized using a metasurface composed of quadratic silicon nanoposts on a fused silica substrate. Transmission and phase responses are simulated for nanoposts of varying sizes to identify geometries that simultaneously yield the desired phase shifts and high transmission efficiency. The pitch remains fixed at 400 nm, while the lateral dimensions of the nanoposts range from 100 to 155 nm. An electron micrograph of the fabricated metasurface is displayed in Fig. 1c. The red box indicates the supercell consisting of 22 unit cells that is repeated to form the whole metasurface. Two identical metasurfaces are fabricated using standard electron beam lithography and etching processes (additional details are provided in the methods section).

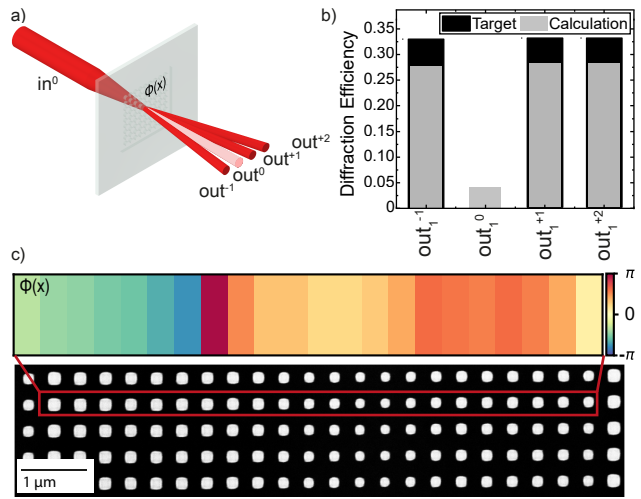

Figure 1: Design of a metasurface-based multi-port beamsplitter. (a) Conceptual sketch of the beam splitting functionality of a single metasurface showing the redistribution of light onto multiple diffraction orders. (b) Comparison between the target and simulated diffraction efficiencies from the designed spatial phase profile. (c) Scanning electron micrograph of the fabricated metasurface. The red box highlights a single supercell with the corresponding calculated phase values shown above.

**Beamsplitting with Classical Light**

In the following, we first characterize the diffraction efficiency of the two metasurfaces individually with classical light. Subsequently, they are cascaded to build an optical network and its interferometric behaviour is analysed. The general concept of the cascaded metasurfaces building the interferometer is illustrated in Fig. 2a. The first metasurface (MS 1) acts as a multiport beamsplitter and the second (MS 2) is used to create the coherent superposition between its inputs. By using two identical lenses, we redirect the light of the output ports of the first metasurface to recombine it at the second metasurface under the same angles. In such a way, different output paths of the first metasurface serve as inputs for the second metasurface and e.g. out$_1^0$ becomes in$_2^0$. Because of the principle of optical reciprocity, cascading the same metasurface twice leads to the recombination into a single beam, provided that the beams impinge on the second metasurface with the correct relative phases. However, experimentally the generated beams have different path lengths, such that the correct phase relation is not achieved. As a result, the new input paths do not recombine into one beam, but every input beam will generate new output paths. We use this behaviour to build an $N \times M$ interferometer. This configuration results in the coherent superposition of multiple diffracted beams in certain output paths, as illustrated in Fig. 2a. The interference between these overlapping diffraction orders redistributes the optical power among these outputs and the relative phase between the inputs allows for active switching serving as a deterministic control parameter for the splitting ratio.

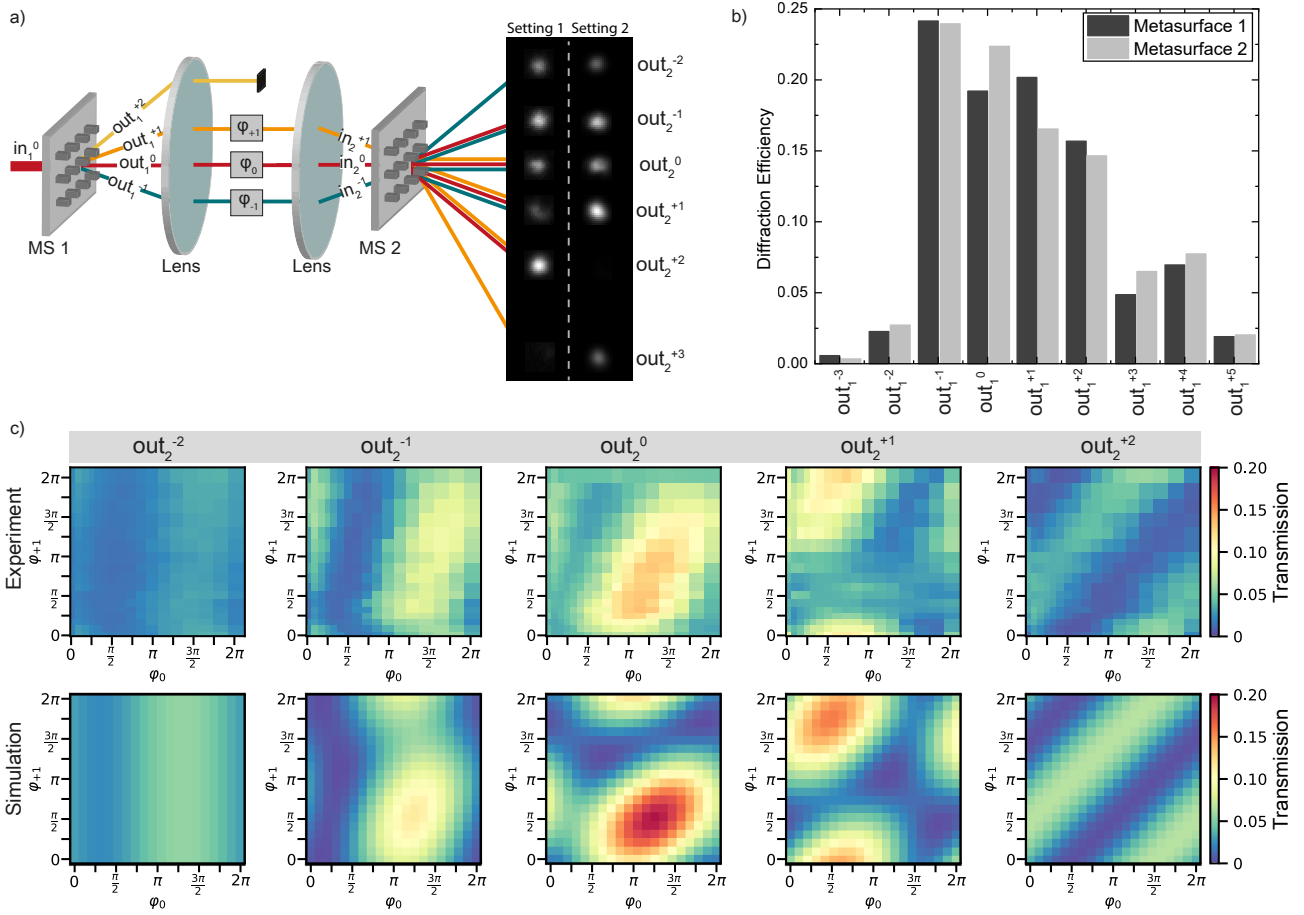

Figure 2: Phase-controlled interference in cascaded metasurfaces. (a) Conceptual illustration of the cascaded metasurfaces. The images show the far-field diffraction spots for two representative phase settings. (b) Comparison of the relative diffraction efficiencies measured for both individual metasurfaces. (c) Experimentally measured (top row) and calculated (bottom row) transmission maps for different phase configurations showing good agreement and confirming the phase-dependent modulation of output intensities due to interference between different input channels.

The measured diffraction efficiencies (relative efficiency) of the individual metasurfaces are displayed in Fig. 2b (the characterization for the other inputs can be found in the Supporting Information). The single metasurface is illuminated with a 785 nm laser diode and the resulting diffraction pattern is recorded using a $4f$-setup and a camera. The intensity of each diffraction order is integrated over a $50 \times 50$ pixel area, background-corrected, and normalised by the total transmitted intensity to obtain the diffraction efficiencies $DE = I_{\text{out}^m} / \sum I_{\text{out}^m}$.

We observe four dominant diffraction spots with diffraction efficiencies from 0.15 to 0.25. Due to fabrication tolerances, part of the light does not accumulate the correct phase and is therefore transmitted in its original direction. Hence, the intensity of output $out_1^0$ is higher than set in the design. Nevertheless, we use this additional beam for our subsequent experiments due to its similar intensity. Next, the phase-dependent output signals of the interferometer are studied. For this, we introduce phase shifters ($\varphi_{+1}$ and $\varphi_0$) between both metasurfaces in the corresponding channels. In the experiment, the channel $out_1^{+2}$ is blocked to simplify the systems.

For each combination of the two phase values, we measure the transmission (absolute efficiency) after the second metasurface for the different output orders. The resulting plots are displayed in the top row of Fig. 2c. The measured transmission maps reveal pronounced modulation of the output intensities as a function of the relative phases. The oscillations are a direct signature of multipath interference. The appearance of each map is directly linked to the number of interfering beams, e.g. the diagonal interference pattern is a consequence that $out_2^{+2}$ depends on both phases $\varphi_0$ and $\varphi_{+1}$. The periodic redistribution of optical power among the output paths demonstrates coherent interference controlled purely by the relative phase. In essence, the cascaded metasurfaces operate as a phase-tunable, high-dimensional interferometer analogous to a network of nested Mach-Zehnder interferometers. This highlights the metasurfaces ability to manipulate the output configurations across multiple modes within a compact footprint.

To predict the phase-dependent transmission, we employ a matrix formalism to calculate the expected intensity for each output path (see Eq. 3). The transmitted intensity is given by $|[E_{out}]|^2 = |[MS_2] \cdot [Phase] \cdot [MS_1] \cdot [E_{in}]|^2$. The vector $E_{in}$ represents the incident field amplitude, which is normalized to unity. The matrix $MS_1$ describes how the incident light is redistributed into the four diffraction orders when entering thorough input path $in_1^0$. It therefore contains four elements corresponding to the transmission coefficients $t_m^{MS1}$ of the (+2,+1,0,-1) diffraction order, listed from top to bottom. Upon diffraction, each beam acquires an additional phase $e^{i\phi_m}$, that is specific to the respective path $m$. The experimentally applied phase shifts are described by the 4×4 phase matrix. The phase modulation is applied to beams propagating in $in_2^{+1}$ and $in_2^0$ and is represented by the phase factors $e^{i\varphi_{+1}}$ and $e^{i\varphi_0}$, respectively. Entries equal to 1 indicate that no external phase shift is applied to path $in_2^{-1}$ and $in_2^{+2}$. When the four beams impinge on the second metasurface, they generate a total of seven output paths $out_2^{+4}$ to $out_2^{-2}$. This transformation is described by the 7×4 matrix $[MS_2]$, which maps the four inputs to the seven outputs. Each incident beam is split according to its corresponding transmission coefficient $t_m$, and each beam acquires an additional diffraction phase $e^{i\phi_m}$.

$$\begin{bmatrix} E_{out_{+4}} \\ E_{out_{+3}} \\ E_{out_{+2}} \\ E_{out_{+1}} \\ E_{out_0} \\ E_{out_{-1}} \\ E_{out_{-2}} \end{bmatrix} = \begin{bmatrix} t_{+2} \cdot e^{i\phi_{+2}} & 0 & 0 & 0 \\ t_{+1} \cdot e^{i\phi_{+1}} & t_{+2} \cdot e^{i\phi_{+2}} & 0 & 0 \\ t_0 & t_{+1} \cdot e^{i\phi_{+1}} & t_{+2} \cdot e^{i\phi_{+2}} & 0 \\ t_{-1} \cdot e^{i\phi_{-1}} & t_0 & t_{+1} \cdot e^{i\phi_{+1}} & t_{+2} \cdot e^{i\phi_{+2}} \\ 0 & t_{-1} \cdot e^{i\phi_{-1}} & t_0 & t_{+1} \cdot e^{i\phi_{+1}} \\ 0 & 0 & t_{-1} \cdot e^{i\phi_{-1}} & t_0 \\ 0 & 0 & 0 & t_{-1} \cdot e^{i\phi_{-1}} \end{bmatrix} \cdot \begin{bmatrix} 1 & 0 & 0 & 0 \\ 0 & e^{i\varphi_{+1}} & 0 & 0 \\ 0 & 0 & e^{i\varphi_0} & 0 \\ 0 & 0 & 0 & 1 \end{bmatrix} \cdot \begin{bmatrix} t_{+2}^{MS1} \cdot e^{i\phi_{+2}} \\ t_{+1}^{MS1} \cdot e^{i\phi_{+1}} \\ t_0^{MS1} \\ t_{-1}^{MS1} \cdot e^{i\phi_{-1}} \end{bmatrix} \cdot [E_{in}]$$

(3)

The matrix model allows several qualitative predictions. Some output paths receive contributions from only a single beam. In these cases, no phase-dependent modulation of the transmission is expected (as for path $out_2^{-2}$). Other output paths are generated by the interference of two or three beams. In such cases, the transmission depends on the relative phase between the contributing paths. For example, the output path $out_2^{-1}$ is generated by the beams propagating through the inputs $out_2^{-1}$ and $in_2^0$. However, only the latter path is phase-modulated. Consequently, the transmission depends solely on the phase $e^{i\varphi_0}$, which results in a vertical modulation pattern in the transmission map. In contrast, the output path $out_2^{\pm 2}$ is formed by interference of beams from $in_2^0$ and $in_2^{+1}$. Since both of these paths are phase-modulated, the transmission depends on both $e^{i\varphi_0}$ and $e^{i\varphi_{+1}}$, leading to the observed diagonal interference pattern.

Figure 2b shows that diffraction orders beyond +2 to -1 are also generated by the metasurface, although with lower efficiency. The experimental data for the output path $out_2^{-2}$ reveal a weak modulation indicating that these additional orders contribute to the interference. Therefore, in the numerical calculations, we include diffraction orders from +2 to -2. Further details are provided in the Supporting Information. The matrices described above capture the general behaviour of the cascaded metasurface system. In the experiment, however, the input path $in_2^{+2}$ is blocked. As a result, the first column of matrix $[MS_2]$ is set to zero. For the calculation, the matrices $[MS_2]$ and $[MS_1]$ are populated with experimentally measured transmission coefficients (see Fig. 2b). Additional information on the derivation are provided in the Supporting Information. Comparing the top and bottom row in Fig. 2c shows a good agreement between the measured and calculated phase maps confirming that the matrix model accurately captures the phase-dependent interference.

**Beamsplitting with Quantum Light**

After characterizing the metasurfaces with the help of classical light, we extend the analysis to quantum light. We used a sample of semiconductor GaAs quantum dots (QDs) grown via molecular beam epitaxy (MBE), with the QDs embedded inside a circular grating resonator (CBR) to enhance the extraction efficiency. Further details on the QD sample can be found in Ref..[31]

The quantum dot sample was cooled to 4 K using a closed-cycle cryostat and excited with a continuous-wave (CW) laser diode at 522 nm (far over-band) to record the photoluminescence spectrum (PL) of the emitter using a spectrometer, which detected the photons using a silicon charge-coupled device (CCD). Once the spectrum was recorded, a bright transition corresponding to a charged exciton was identified and selectively excited using a near-resonant pulsed laser (see Fig. 3b) in a phonon-assisted scheme[32] while allowing a small amount of the CW laser to neutralize the charge environment and allow resonant excitation.

To perform correlation measurements, the pulsed laser was filtered using band-rejection filters, followed by a transmission monochromator (TM), which isolated the T transition line from the QD. We first characterize the single photon emission from the quantum dot sample by measuring the second-order intensity correlation function of the generated light with a Hanbury Brown and Twiss setup. After the TM, the light is coupled into a 50:50 fibre-based beamsplitter (BS) and sent to two superconducting nanowire single-photon detectors (SNSPDs, Single Quantum). The coincidence counts between the two outputs of the BS are recorded for different delays $\tau$. The second-order intensity $g^{(2)}(\tau)$ shown in Figure 3b reveals nearly background free single photon emission. The measured degree of second order coherence $g^{(2)}(0) = 0.063 \pm 0.003$ confirms strong anti-bunching, demonstrating that the emission is dominated by true single photon events that cannot arise from classical light sources. Next, we replace the fibre-based BS with the single metasurface to perform both second- and third-order correlation functions. The metasurface acts a multiport beamsplitter that divides the incident single photon wavefront into mainly four distinct spatial modes. This enables simultaneous access to multiple correlation orders using a single compact optical element. The four output paths of the metasurface are individually coupled into single mode fibres connected to four channels of the SNSPDs. Figure 3c shows two representative second-order intensity correlations measured between output pairs (0,+1) and (0,-1). Both measurements show nearly background-free anti-bunching with a degree of second order coherence of $g^{(2)}(0) = 0.034 \pm 0.003$ and $g^{(2)}(0) = 0.036 \pm 0.003$, confirming that the metasurface preserves the quantum nature and coherence of single photon emission.

We then extend the analysis to third order intensity correlations $g^{(3)}(\tau_1, \tau_2)$. Two measurements are shown in Fig. 3c for the output path combinations (0,+1,-1) and (0,+1,+2). The data exhibits strongly suppressed multi-photon emission probabilities, visible as three characteristic lines of reduced coincidence counts: i) the first (0) and second (+1) photon arrive at the same time (the central column), ii) the first (0) and third (-1) photon arrive at the same time (the central row) and iii) the second (+1) and third (-1) photon arrive simultaneously (the diagonal line). These features are the direct quantum signatures of photon anti-bunching in a three-photon coincidence measurement. Unlike conventional bulk or fibre-based beamsplitter the metasurface allows simultaneous access to multi-path photon correlations and provides a scalable approach for integrated quantum-optical characterization. In essence, the metasurface implements a spatially multiplexed quantum beam splitter in a single planar component, simplifying the architecture for higher-order quantum correlation and interference measurements. Building on these results, we investigate the interference of single photons using the cascaded metasurfaces. The first metasurface splits the incident light into the four distinct orders. Two of these outputs are selected and with the help of two lenses directed onto the second metasurface, while the remaining paths are blocked to simplify the system. A phase shifter placed in one of the path introduces a tunable relative phase $\varphi_{+1}$ before they are recombined on the second metasurface. The resulting spatial intensity distribution is recorded with a CCD, from which the diffraction efficiencies of the individual output paths are extracted. Figure 3d shows the diffraction efficiency per output path as a function of the applied relative phase shift. The intensity in the output paths $out_2^0$, $out_2^{+1}$, and $out_2^{+2}$ oscillates sinusoidally with the phase providing evidence for phase-dependent redistribution of the photon probability amplitude among the output channels. Although only a single photon is present in the interferometer at a time, as demonstrated by $g^2(0)$, the interference persists, demonstrating that the photon's wavefunction simultaneously occupies both optical paths. At the second metasurface constructive and destructive interference depending on the relative phase occur, leading to the observed modulation of the diffraction efficiencies.

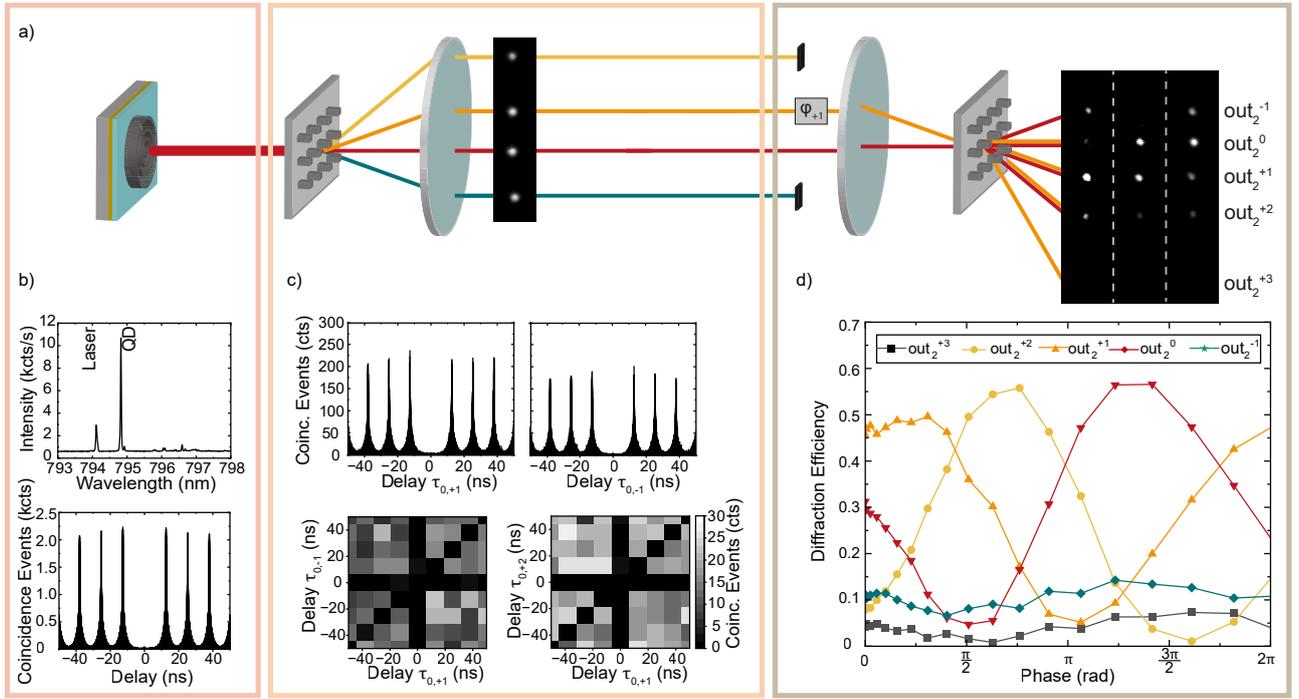

Figure 3: Quantum measurements on a metasurface-based interferometer. (a) Schematic drawing of the cascaded metasurface with charge-coupled device images showing the phase dependent interference. (b) Quantum dot emission spectrum and second-order correlation measurement confirming single-photon emission through anti-bunching. The spectrum was filtered with only notches. (c) Measured second- and third-order correlations showing that the metasurface maintains quantum coherence and non-classical light correlations. (d) Phase-controlled redistribution of the light revealing coherent interference at the single-photon level.

The cascaded metasurfaces form a metasurface-based interferometer analogous to a Mach-Zehnder configuration, however, with extended dimensionality. This architecture highlights the potential of metasurfaces to realize multiport quantum interference networks with specifically engineered properties.

## Conclusion

We have built a metasurface-based interferometer from two cascaded dielectric metasurfaces, each functioning as a multiport beamsplitter and show that it is capable of coherent multiport interference using both classical and quantum light. First, the diffraction efficiencies of the individual metasurfaces were characterized with classical light. Quantum correlation measurements confirm that the metasurface preserves the non-classical correlations of the incident quantum light, demonstrating its compatibility with quantum optical applications. By cascading two nominally identical metasurfaces, we realized a high-dimensional interferometer in which the interference between the outputs can be deterministically tuned by adjusting the relative phase between the inputs. The resulting oscillatory transmission was observed for classical and quantum light and can be predicted by the developed matrix model.The ability to control the diffraction efficiencies and output distribution through phase tuning highlights the potential of metasurfaces for reconfigurable optical routing and switching. The ability to control the diffraction efficiency and output distribution through phase tuning highlights the potential of metasurfaces for reconfigurable optical routing and switching. In the present work, the metasurface phase gradient was designed along a single spatial direction. However, the concept can be naturally extended to two dimensions by simply implementing a two-dimensional spatial phase profile. In such a design, the metasurface supercell would differ in the *x*- and *y*-directions, enabling additional degrees of freedom for tailoring the diffraction pathways. This design flexibility illustrates the broader potential of metasurfaces for implementing complex multipath optical transformations. Recent studies have demonstrated that metasurfaces can be directly fabricated on the end facets of optical fibers, enabling wavefront shaping and beam steering within fiber-based platforms.[33,34] Integrating metasurfaces with optical fibers could therefore allow the realization of compact beamsplitting devices in which beam splitting occur directly at the fiber interface. Overall, our results establish cascaded metasurfaces as a versatile and compact platform for implementing high-dimensional interferometers in both classical and quantum regimes.

## Acknowledgements

This work is supported by the Deutsche Forschungsgemeinschaft (German Research Foundation) through the transregional collaborative research center TRR142/3-2022 Project A08 (231447078). The project was supported by the German Federal Ministry of Research, Technology and Space (BMFTR) through the projects QR.N (16KIS2206) and QPIC-1 (13N15856). This project has received funding from the European Research Council (ERC) under the European Union's Horizon 2020 research and innovation program (LiNQs, grant agreement 101042672). R.T and M. B. R. acknowledge support from MUR (Ministero dell'Universita e della Ricerca) through the PNRR MUR project PE0000023-NQSTI. THL acknowledges support from the German Federal Ministry of Research, Technology and Space within the project Qecs (FKZ: 13N16272).
We would like to thank Silke Kuhn for help with the preparation of the quantum dot sample.

## Supporting Information

The supporting information contains details on the metasurface fabrication, simulated transmission coefficients and phase values for different nanopost sizes, further information on the experimental design, and the derivation of the metasurface matrix.

# Supporting Information: Cascaded Metasurface Interferometer for Multipath Interference with Classical and Quantum Light

**Fabrication**

The metasurface fabrication requires a multi-step process starting with the deposition of 600 nm amorphous silicon on a fused silica substrate by plasma enhanced chemical vapor deposition. Prior to the deposition the fused silica substrates are cleaned in acetone, isopropanol, and distilled water for 10 min each at 30 °C. Additionally, cleaning in an oxygen plasma is performed for 10 min to remove any organic residues. A 160 nm thick polymethyl-methacrylate (PMMA) resist layer is spin-coated onto the substrate and baked at 165 °C for two minutes for hardening. The pattern is transferred to the resist during electron beam lithography. A Raith Voyager system with an acceleration voltage of 50 kV and an aperture size of 30 µm is used. After developing the samples in MIBK, a 20 nm thick chromium mask is deposited using electron beam evaporation. After the liftoff in hot acetone and DMS, the pattern is transferred from the resist to the Cr mask. Subsequently, the silicon is etched with inductively coupled plasma reactive ion etching with $SF_6$ and $C_4F_8$ as the process gases. Lastly, the remaining Cr mask is removed by wet etching.

**Transmission Coefficients of Metasurfaces**

To design the metasurface, full-wave electromagnetic simulations were performed to determine the transmission amplitude and phase as a function of nanopost size for a fixed incident wavelength of 785 nm. The nanoposts have a rectangular cross section with side lengths ranging from 100 nm to 155 nm. This parameter range was selected because it allows continuous phase coverage while maintaining high transmission efficiency. Figure S1 shows the simulated transmission phase and amplitude for x- and y-polarized incident light. The upper panels show the phase response of the nanoposts, while the lower panels show the corresponding transmission amplitudes. The region used for the metasurface design corresponds to the lower-left corner of the plots. In this regime the transmission amplitude remains close to unity, which minimizes losses while still enabling the required phase control. Sharp variations in the transmission amplitude correspond to resonant behaviour of the nanoposts. To ensure stable device operation, the metasurface design avoids operating exactly at these resonances.

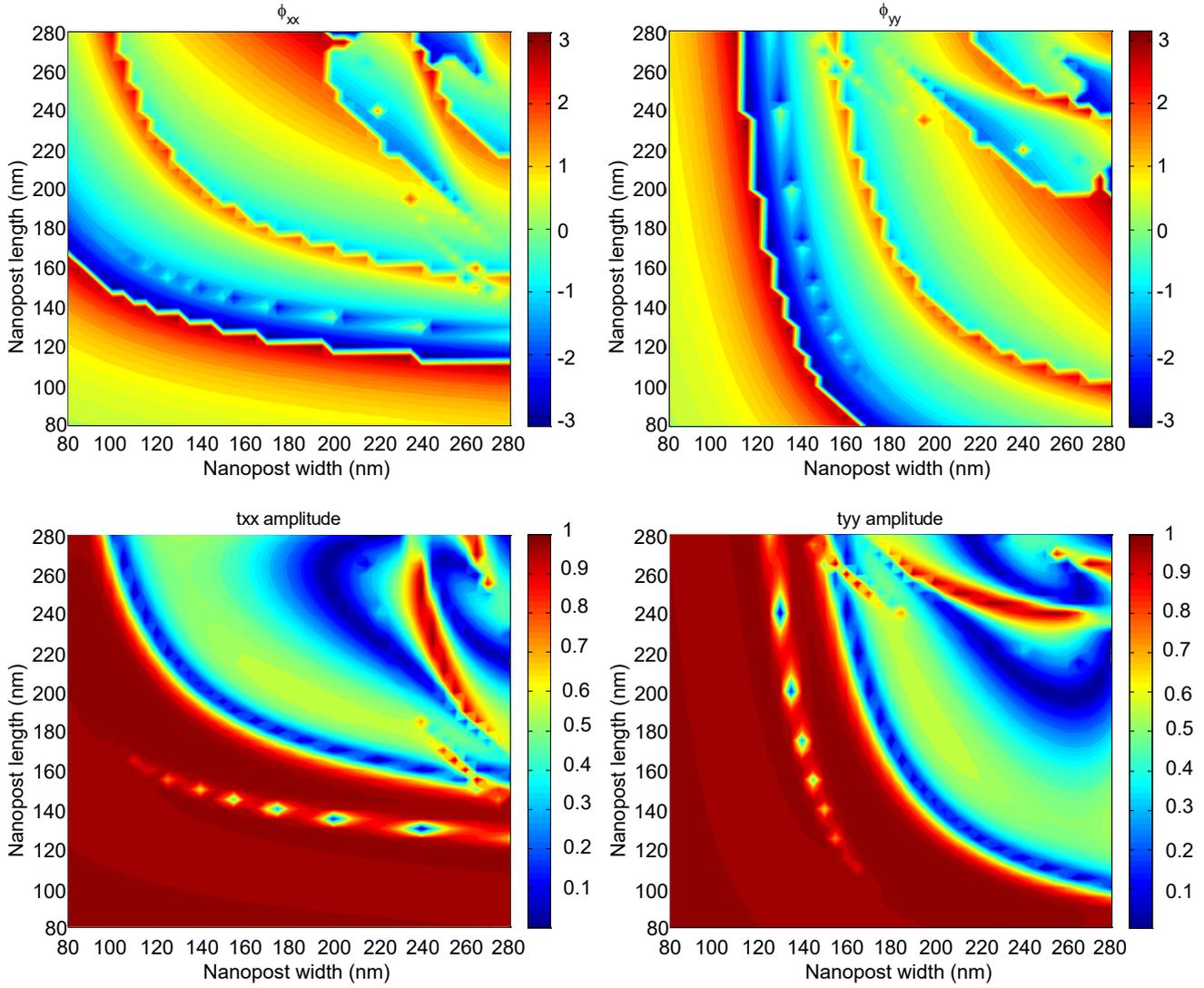

Figure S1: Simulated phase and transmission amplitude of the silicon nanoposts at a wavelength of 785 nm. The upper panels show the achievable phase shift, while the lower panels show the corresponding transmission amplitude. The metasurface design utilizes nanopost dimensions in the range of 100 nm to 155 nm, where the transmission amplitude remains close to unity.

The sizes of nanopost in the design are:

| Antenna index | 1 | 2 | 3 | 4 | 5 | 6 | 7 | 8 |
|---|---|---|---|---|---|---|---|---|
| Size (nm) | 150×155 | 150×150 | 140×145 | 140×145 | 140×145 | 140×140 | 140×140 | 135×135 |
| Antenna index | 9 | 10 | 11 | 12 | 13 | 14 | 15 | 16 |
| Size (nm) | 120×125 | 110×115 | 110×115 | 100×105 | 100×100 | 105×110 | 115×120 | 125×125 |
| Antenna index | 17 | 18 | 19 | 20 | 21 | 22 | | |
| Size (nm) | 125×125 | 125×125 | 125×125 | 125×125 | 115×120 | 155×155 | | |

**Further Details on the Experimental Design**

In the following the experimental setup for the classical measurements is described. A laser diode with a central wavelength of 785 nm is used as a coherent light source. The polarization is set to horizontal with the combination of a linear polarizer and a half-wave plate. The light is focused down to the size of the metasurface with a 300 mm lens. At the first metasurface (MS1) the incident light is split into the diffraction orders. The combination of two lenses (f=35 mm) is used to direct the light on the second metasurface (MS2). Between these two lenses, the phase is modulated. This is achieved by rotating 150 µm thin glass slide on a rotation stage in increments of 0.5°. At the second metasurface the incident light is diffracted again and the resulting paths are collected with an microscope objective MO (NA 0.7, 20x). The back focal plane (BFP) of the MO is imaged with a 4f-setup consisting of two lenses onto the CCD chip. If the first lens of the 4f-setup is removed the real space is imaged onto the CCD meaning that the metasurface is visible on the camera.

For the correlation measurements with quantum light, the MO, the two lenses and the CCD are exchanged. The photons in the different output paths are collected with fiber couplers and sent to the single photon detectors.

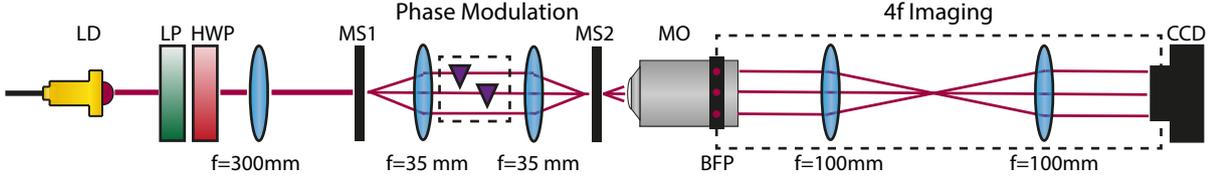

Figure S2: Schematics of the experimental setup for measurements with classical light.

### Transfer Matrix

To evaluate the performance of the metasurface interferometer, we experimentally characterized the diffraction efficiencies (DE) for different input ports. In particular, the diffraction efficiency of a single metasurface was measured not only for normal incidence (corresponding to the 0th diffraction order) but also for beams incident under the angles corresponding to the first, second, and minus first diffraction orders. The measured diffraction efficiencies are summarized in Table 1. Each column corresponds to a different input port, while each row corresponds to an output diffraction order.

| Output Path | DE for input in -1 | DE for input in 0 | DE for input in +1 | DE for input in +2 |
|---|---|---|---|---|
| 4  | 0.039 | 0.0461 | 0.121 | 0.171 |
| 3  | 0.018 | 0.068  | 0.072 | 0.201 |
| 2  | 0.022 | 0.153  | 0.169 | 0.230 |
| 1  | 0.201 | 0.217  | 0.202 | 0.201 |
| 0  | 0.200 | 0.170  | 0.280 | 0.024 |
| -1 | 0.202 | 0.269  | 0.024 | 0.003 |
| -2 | 0.234 | 0.024  | 0.004 | 0.012 |

These measurements confirm that the metasurface primarily redistributes optical power among the designed diffraction orders, while higher orders remain relatively weak.

### Further Information on Matrix Model

In the main text we present the matrix, that describes the general and ideal case for the transformation that the second metasurface given as:

$$\begin{bmatrix} t_{+2} \cdot e^{i\phi_{+2}} & 0 & 0 & 0 \\ t_{+1} \cdot e^{i\phi_{+1}} & t_{+2} \cdot e^{i\phi_{+2}} & 0 & 0 \\ t_0 & t_{+1} \cdot e^{i\phi_{+1}} & t_{+2} \cdot e^{i\phi_{+2}} & 0 \\ t_{-1} \cdot e^{i\phi_{-1}} & t_0 & t_{+1} \cdot e^{i\phi_{+1}} & t_{+2} \cdot e^{i\phi_{+2}} \\ 0 & t_{-1} \cdot e^{i\phi_{-1}} & t_0 & t_{+1} \cdot e^{i\phi_{+1}} \\ 0 & 0 & t_{-1} \cdot e^{i\phi_{-1}} & t_0 \\ 0 & 0 & 0 & t_{-1} \cdot e^{i\phi_{-1}} \end{bmatrix} \quad (1)$$

However, the experimental data shows that a more complex description is necessary. In the following and with the help of Figure S3 the derivation of the final matrix is described. The most simple way to describe the nominal transfer matrix is as follows (Fig. S3a):

$$\begin{bmatrix} t_{+2} & 0 & 0 & 0 \\ t_{+1} & t_{+2} & 0 & 0 \\ t_0 & t_{+1} & t_{+2} & 0 \\ t_{-1} & t_0 & t_{+1} & t_{+2} \\ 0 & t_{-1} & t_0 & t_{+1} \\ 0 & 0 & t_{-1} & t_0 \\ 0 & 0 & 0 & t_{-1} \end{bmatrix} \tag{2}$$

This matrix describes the case where the each input order generates the four designed outputs paths. The first column describes how a beam incident in $in_2^{+2}$ is split again into four paths $out_2^{+4}$ to $out_1^0$. The second column describes how a beam incident in $in_2^{+1}$ is split again into four paths $out_2^{+3}$ to $out_1^{-1}$ and so on. If all four inputs are incident at a second metasurface six output beams (the +4 to -2 order) are generated and no phase-dependent behaviour is observed for output $out_2^{-2}$ as this output path is only generated by a single beam. This description, however does not capture the experimental data well. The agreement between calculation and data can be further enhanced if the following matrix is considered:

$$\begin{bmatrix} t_{+2} & 0 & 0 & 0 \\ t_{+1} & t_{+2} & 0 & 0 \\ t_0 & t_{+1} & t_{+2} & 0 \\ t_{-1} & t_0 & t_{+1} & t_{+2} \\ t_{-2} & t_{-1} & t_0 & t_{+1} \\ 0 & t_{-2} & t_{-1} & t_0 \\ 0 & 0 & t_{-2} & t_{-1} \end{bmatrix} \tag{3}$$

This matrix accounts for the circumstance that the metasurface also generates higher orders (as described in the main text). In this case we also consider that the metasurface also generates the -2.order for every input beam. As indicated in Fig.S3b a phase-dependent transmission map is now expected for the output $out_2^{-2}$. Now a last difference between calculation and data remains, that is that in the calculation path $out_2^0$ and $out_2^{+1}$ display the same pattern in the phase map. This is in agreement with the calculation because in $out_2^0$ and $out_2^{+1}$ the same beams interfere with each other. However, the experimental data shows the same pattern but phase-shifted. We account for this by introducing the additional phase terms $e^{i\phi_m}$ which describes that every diffraction order acquires a specific additional phase term upon deflection. With this phase the different behaviour of $out_2^0$ and $out_2^{+1}$ can be explained. For the calculation of the presented phase maps in the main text the following matrix was used (Fig. S3c):

$$\begin{bmatrix} 0 & 0 & 0 & 0 \\ 0 & t_{+2} \cdot e^{i\phi_{+2}} & 0 & 0 \\ 0 & t_{+1} \cdot e^{i\phi_{+1}} & t_{+2} \cdot e^{i\phi_{+2}} & 0 \\ 0 & t_0 & t_{+1} \cdot e^{i\phi_{+1}} & t_{+2} \cdot e^{i\phi_{+2}} \\ 0 & t_{-1} \cdot e^{i\phi_{-1}} & t_0 & t_{+1} \cdot e^{i\phi_{+1}} \\ 0 & t_{-2} \cdot e^{i\phi_{-2}} & t_{-1} \cdot e^{i\phi_{-1}} & t_0 \\ 0 & 0 & t_{-2} \cdot e^{i\phi_{-2}} & t_{-1} \cdot e^{i\phi_{-1}} \end{bmatrix} \tag{4}$$

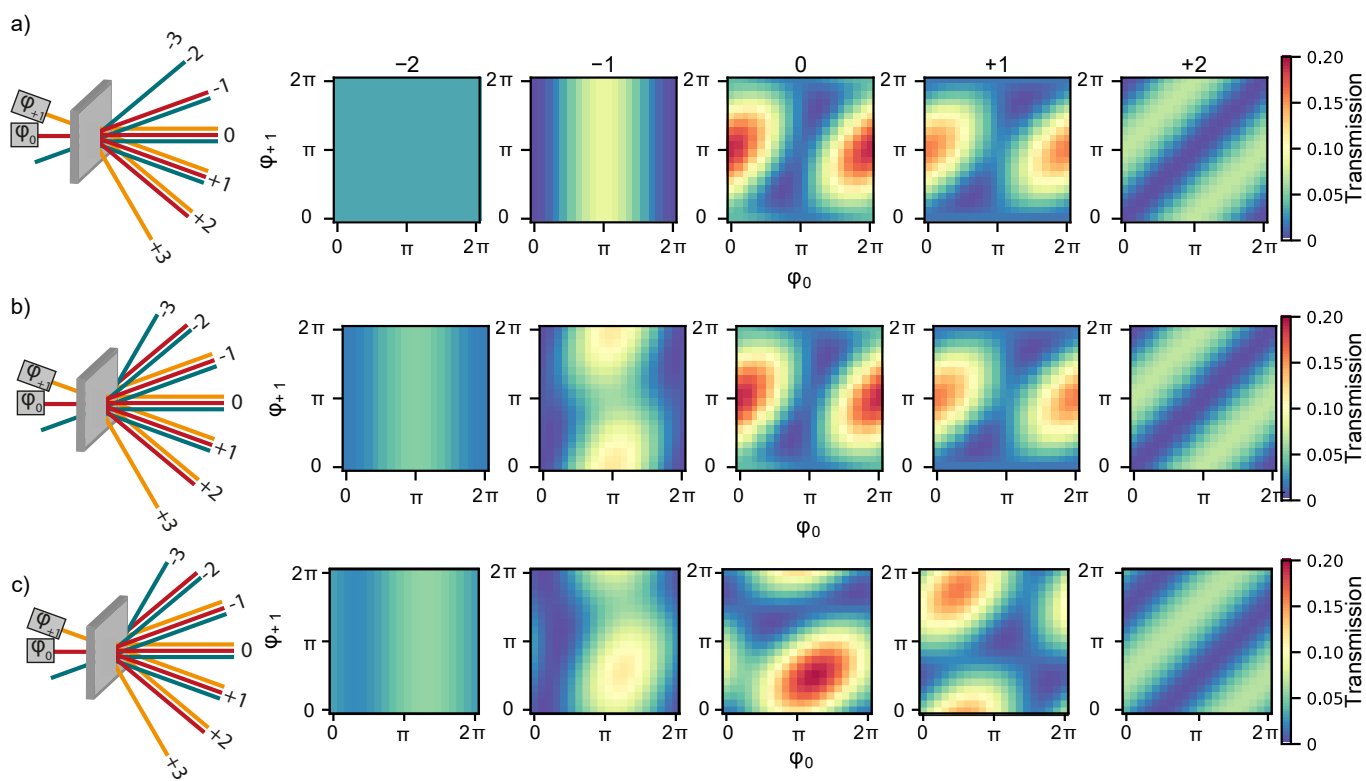

Figure S3: The calculated phase maps differ for different transfer matrices.